\documentclass{article}
\usepackage{graphicx} 
\usepackage{fullpage}
\usepackage{hyperref}
   \hypersetup{
           breaklinks=true,   
           colorlinks=false,   
        }
\usepackage{changepage}
\usepackage{authblk}

\usepackage[sort&compress]{natbib}
\setcitestyle{numbers}
\setcitestyle{square}
\begin{document}

\title{AI Procurement Checklists: Revisiting Implementation in the Age of AI Governance}
\author[1]{Tom Zick\thanks{tzick@cyber.harvard.edu}}
\author[1]{Mason Kortz\thanks{mkortz@cyber.harvard.edu}}
\author[2]{David Eaves\thanks{d.eaves@ucl.ac.uk}}
\author[1]{Finale Doshi-Velez \thanks{finale@seas.harvard.edu}}
\affil[1]{Harvard University}
\affil[2]{University College London}

\maketitle

Public sector use of AI has been quietly on the rise for the past decade, somewhat lethargically trailed by efforts to regulate it. Failed attempts at techno-utopias, like the Google-funded Sidewalk Labs Toronto Waterfront experiment~{\citep{The_Guardian_2020}}, began to foreground the potential harms of public AI. But it was only recently, with the advent of ChatGPT and passage of the EU AI Act, that efforts to regulate government adoption of automated tools has broken into the zeitgeist. At the state and local level,  Vermont publicly released an inventory of all their AI tools~\citep{AIRegula99:online}, while Connecticut signed into law a bill that would require additional procurement scrutiny for AI-based products~\citep{Connecti65:online}. California recectly introduced comprehensive  legislation to govern their internal AI use~\citep{Checking31:online},  and other major states are following suit~\citep{2024AISt34:online}.  At the Federal level,  a significant portion of the Executive Order on Artificial Intelligence released late last year centers on responsibly increasing government AI use \citep{Executiv42:online}. Collectively, these efforts have centered responsible adoption of algorithmic systems with reporting frameworks for general effectiveness, privacy implications, and  potential for discriminatory bias. 

While simple to articulate, promoting ethical and effective roll outs of AI systems in government is a notoriously elusive task. On the one hand there are hard-to-address pitfalls associated with AI-based tools, including concerns about bias towards marginalized communities, safety, and gameability. On the other, there is pressure not to make it too difficult to adopt AI, especially in the public sector which typically has fewer resources than the private sector––conserving scarce government resources is often the draw of using AI-based tools in the first place. These tensions create a real risk that procedures built to ensure marginalized groups are not hurt by government use of AI will, in practice, be performative and ineffective~\citep{DataSoci11:online}.

To begin to address these issues in the US, we look to jurisdictions with more mature regulations around government AI use. Countries like Brazil, Singapore and Canada have been grappling with how to navigate the tensions associated with rolling out public sector AI systems for a number of years. Collectively, these governments have implemented risk categories, disclosure requirements and assessments into the way they procure AI tools. Core to these efforts is the  AI procurement checklist~\citep{checklists}, a type of impact assessment that government agencies must conduct before acquiring certain AI-based tools. 

Procurement is a natural focal point for managing government AI deployment. For one, it is ubiquitous. Procurement is the process by which governments acquire new tools, from cloud computing services to pencils and everything in between. In theory, any AI-based tool a government might use will go through some type of procurement process. Additionally, procurement is fundamentally organized around ensuring governments acquire the best products for furthering their policy goals. This means that government procurement already incorporates processes meant to ensure agencies define clear metrics for success, solicit products or services that comply with these metrics, and use them to evaluate competing vendors. AI procurement checklists augment these existing processes. On the one hand, checklists aim to assist in the selection of objectively good products that will efficiently achieve legitimate government ends without over- or under-investing in new technologies. At the same time, AI checklists put a strong emphasis on responsible and ethical use of AI, which requires both objective protocols and value judgments that amount to policy determinations. 

In this work, we investigate two implemented checklists: the Canadian Directive on Automated Decision-Making (CDADM)~\citep{CADM} and the World Economic Forum’s AI Procurement in a Box (WEF)~\citep{AIProcur11:online}. While CDADM is a jurisdiction specific initiative, the WEF checklist was developed as a mold that various jurisdictions have been able to adopt. We discussed these checklists with government officials who had seen them at work in the Canadian, Brazilian, and Singaporean governments. Our own inspections and our conversations resulted in three major observations that can inform the current wave of regulatory interest:

\begin{itemize}

\item \textbf{Technical assessments require an AI expert to complete---and we don’t have enough experts.} Technical checklists are most useful when they operate as reminders for experts. Pilots exhibit safer flying when they follow technical checklists. That does not mean that a lay person can use those same checklists to fly the plane. Similarly, the knowledge of a generalist civil servant even when augmented by an AI procurement checklist will simply not be enough.  This is unlikely to change in the foreseeable future. We need to think carefully about how we will create enough expertise and who will have access to this expertise.

\item \textbf{When defining ”AI,” it is very difficult to sweep in all systems that need additional oversight---we need to close the loopholes.}  Some jurisdictions do not require procurement checklists for systems developed in-house, or ones that cost less than a certain amount.  Given how broad the impact of AIs can be, it is important to close these loopholes; all AIs should require a checklist process, even if it is simply to ascertain that the system is of sufficiently low risk.

\item \textbf{No expert audit will be perfect. Public transparency is a necessary component for deploying effective and ethical AI systems.} Not only are AI systems complex, but their impacts can also be complex and heterogeneous.  Even a very good expert may not be able to catch all the errors. Currently, certain types of AIs can be procured without scrutiny, and the results of checklist assessments may or may not be made public.  Especially as third-party experts will almost certainly be needed to complete the checklists properly, transparency is needed to establish norms about how guidelines should be interpreted and catch aspects that the experts missed. Algorithmic registries on their own are not sufficient.

\end{itemize}

All of the above are in tension with making AIs easy to procure. There is a real pressure within governments to the avoid additional overhead that comes with enhanced oversight. Without deep attention to the experience of jurisdictions with more mature public AI governance, emerging efforts within the United States are unlikely to successfully navigate the fine line that is responsible government AI deployment. Throughout this work, we elaborate on these tensions and takeaways, drawing on the experience of civil society members and government officials as well as our own analysis.

\section{Technical assessments require an AI expert to complete ---\\and we don’t have enough experts}

Governments like Canada and Brazil have implemented exhaustive technical checklists into government procurement of AI systems. Some parts of these checklists will be familiar to anyone with experience in procurement: best practices for requesting and assessing bids; involving diverse, multidisciplinary teams spanning all stakeholders and areas of expertise; and ascertaining relevant legal requirements. Others, such as those related to training data and model implementation, or privacy and security implications, all  require technical expertise to evaluate. Just as a flight safety checklist must be performed by a pilot or flight engineer to be effective, these AI assessments require technical experts—and this is unlikely to change in the foreseeable future.
\newpage

\noindent For example, the WEF guidelines include instructions to:\\

\begin{adjustwidth}{2cm}{}
“...enable end-to-end auditability with a process log that gathers the data across the modelling, training, testing, verifying and implementation phases of the project life cycle.” \\ 
\end{adjustwidth}

\noindent Similarly, the CDADM includes requirements that systems be “tested for unintended data biases.” \\

But what exactly should be logged?  What is a sufficient description of the training data?  Suppose a vendor submits a report that demonstrates that the system has high accuracy across a list of protected categories (e.g. race, gender, age), does that mean it has no unintended biases?  To that last question: consider a system that predicts recidivism that has the same error rate for white and black people, but when it gets things wrong for black people, it miscategorizes them as high risk, whereas it miscategorizes white people as low risk. The accuracies are the same, but the system is discriminatory.  These questions are nuanced.  Without AI experts to probe deeply at these requirements, many government assessments may end up turning on whether a vendor offered a plausible answer to questions evaluating potential harms, rather than a satisfactory answer. 

\paragraph{Red-Teaming Exercise: Exposing Where Experts are Needed.}

Our general take-away was that while current checklists provide value—in particular, we found only a few blindspots—they should be interpreted not as creating a process that anyone can follow, but a checklist to ensure that an expert does not forget to check some category of concerns.  For example, a housing inspector has a list to check various parts of a house during an inspection, but a lay person may not be able to use that list effectively.  A related value of a checklist is that it emphasizes to procurement agent to know what they don’t know, and thus reach out for appropriate engagement with experts (in house or outside) who do.

We enlisted the Harvard Data to Actionable Knowledge Lab to run a redteaming exercises on the WEF guidelines, where they brainstormed ways to subvert the intention behind the checklist while still complying with the text. Broadly, most subversions involved a vendor or developer choosing to interpret requirements narrowly or formalistically to suit their interests.  

For example, a vendor might claim that an algorithm for scoring cancer severity or recidivism risk is too complex to explain (to avoid a transparency requirement).  However, in many applications–including those in health and criminal justice–modern AI methods can be trained to identify simple, easily inspected models that have the same accuracy as complex, deep models.  And in situations when such simple models are not sufficient, modern interpretation techniques can still help peek inside a deep model black box. A procurement agent without sufficient knowledge of AI systems may take the vendor’s claim at face-value, and not press further. 

As another example, a vendor may provide a set of summary statistics of demographics of the population on which an algorithm was trained—say the age, gender, race, and insurance status of patients used to train a predictor for prioritizing a vaccine roll-out.  A procurement agent may check to make sure that their intended population for the use of the algorithm falls within these demographics.  However, they may not know enough to ask for other key statistics, such as the proportions of patients which were immunocompromised or were front-line workers. An expert would have the tools to identify the absence of important details and know how to probe further for relevant information. Through our red-teaming exercise, we highlight three areas in which expert involvement seemed especially critical:\\  

\textit{Identifying Data Issues.}  Both the Canadian Directive and the WEF guidelines recognize the importance of data quality to AI systems.  It is well understood that AI systems can easily replicate the biases in their data.  Certain basic checks, such as how accurately the system performs across different subgroups, can be performed with limited statistical expertise.  However, determining whether that level of reporting is sufficient - and what additional measures to request if it is not - requires expert review.  In some cases, such expertise may be available in-house; certain agencies we engaged with had very deep expertise.  However, in other cases, units may have none at all and require external experts.

For example, suppose that the accuracy of a new algorithm is high across all protected or sensitive categories (e.g. race, gender, or age). A vendor might put this forth as evidence that “the data... are tested for unintended data biases,” as required by the Canadian Directive. To a government agent who is an expert in procurement but a novice in machine learning, this might make intuitive sense. However, a machine learning expert would know to investigate the errors more closely to determine the valence of the error. Are there more false negatives? False positives? Then, in conjunction with a domain expert, they can determine whether those errors appeared to be random or due to some unobserved confounding variable (like income or neighborhood).

More importantly, an expert would recognize that high performance does not mean low bias.  For example, consider an algorithm that uses healthcare utilization as a proxy for severity, and allocates resources based on predicted severity of illness within a subpopulation.  The  algorithm might accurately predict healthcare utilization across race, gender, etc.  However, it would not understand that utilization depends not just on illness severity but also on social barriers to access.  As a result, the algorithm will under-allocate resources to populations with prior limited access (since they had lower prior utilization). This will further limit future access, entrenching disparities even while vindicating the algorithm’s “accuracy.”  In this way, a seemingly innocent simplification for model training has direct consequences for healthcare availability.  An expert would also know the host of other steps—ranging from how missing data are handled to what procedures are used for data collection—that might impact the quality of the final algorithm.\\

\textit{Post-procurement Checks.}  Both of the checklists we reviewed recommend ongoing, post-procurement monitoring. However, neither prescribes specific processes, metrics, or schedules for such monitoring.   Similar to checking for data quality, an expert can assist in setting up an auditing regime that will provide adequate review after an AI system is in operation.

In terms of systems in operation, both sets of guidelines, and in our discussion with experts, there was clear recognition that certain checks must be performed if the AI model is updated.  Less explicit in the guidelines was that it is equally important for experts to monitor for shifts in the data.  For example, suppose a new guideline is released reducing the recommended age for cancer screening.  Suddenly, an algorithm that assists with tumor detection will be receiving scans of younger patients whose images may have different properties than the older patients it was trained on.  Similarly, major shocks—for example, the Covid-19 pandemic—may change everything from how people interact with transit, to how they purchase goods, or use the internet.  

Currently, machine learning algorithms tend to perform poorly on data that is significantly different from the data on which they were trained. Even knowledge transfer between related problems can be unreliable at best.  For example, an algorithm trained to predict movie ratings during an era of mail-in DVDs (when people often watched the entire movie) may no longer apply–for the same users, for the same movies–once movies start being streamed (and people are more willing to switch to a new movie part-way through).  Thus, it is important to check the performance of an algorithm not only when it is updated, but also if the data being analyzed are sufficiently different from those on which the algorithm was trained. Identifying the types of shifts that require an in-depth review itself requires expert input.

A related concern is one of feedback loops: the introduction of the algorithm itself may change the data itself.  For example, if an AI-based alert system is integrated into a hospital setting, and clinicians start using it, then the new data being produced will look very different from the data under which the algorithm was trained.  Similarly, an explainable essay grading system may shift the way in which students write essays to get a high score.  A computational expert working with a domain expert can determine what kinds of feedback loops to monitor for.\\

\textit{Human Validation of AI Outputs.}  Finally, both guidelines note the importance of having a human in the loop when making high risk decisions.  Implicit in this guideline is that as long as a human is making the final decision, then they will do so properly.  However, a host of human factors literature shows that this is not the case: people tend to over-rely on algorithmic recommendations.  Sometimes this is because the system appears to be highly intelligent; sometimes this is simply because it avoids cognitive labor. Designing a system where one can reasonably claim the human is making the final decision is not straightforward.  Current guidelines require that operators of an AI system receive appropriate training but without a human factors expert it will be difficult to ensure a system is designed for shared decision making. Moreover, it will be difficult to monitor whether this shared decision making is even happening.

\paragraph{Finding and Integrating AI Experts into Procurement.}
The Biden administration's Executive Order on the Safe, Secure, and Trustworthy Development and Use of Artificial Intelligence highlights that the federal government needs to recruit AI talent. Our conversations with first movers in the regulatory space indicate that is a necessary component—we need more experts—and to be effective, these experts also need to be appropriately integrated into the AI procurement process. Some of the governments we spoke with had existing agencies with technical capacity and had designated units within them to assist with technical interpretation. The most sophisticated of these would sometimes even create tools that could be used by many government agencies to evaluate AI products.

That said, \textit{how} experts are integrated into government can completely determine their efficacy. Even when such expert units were available, the officials we spoke with flagged that getting government agencies to seek help in procurement was no easy task. They worked at making their office approachable, encouraging potential procurers of AI to come and discuss, early on, what would make AI implementation successful. For the moment, the rate of AI procurement has been fairly low—a few systems per year—and these offices have been able to invest in outreach and approachability. However, this approach will not scale as AI systems become more commonly used. It would take a sizeable investment, authorized by Congress, to scale the number of experts to meet the Biden administration’s ambitious AI goals.

It is also worth noting that—at least in the case of the Canadian and Singaporean experts we spoke to—there was a strong desire to stay connected to external experts as well as to iterate and improve internal processes. This included making presentations about checklist application to stakeholders within and outside government, such as academia and civil society.  This is precisely the type of behavior that is needed in these early days when the risks and challenges are less clearly understood and the impact on public trust could be significant. It is imperative that both bureaucratic and political leaders do not impede this important sharing.

Finally, we highlight that even with the recognition that experts are needed for certain parts of the checklists, there is still the question of how to source the experts.  That is, who verifies that the expert has sufficient expertise?  As the use of AI checklists becomes more common, we imagine that a governing body will be needed to certify AI experts, similar to other inspection and audit related disciplines.

\section{Procurement Loopholes Exist}

Though procurement is a process that touches nearly all government functions, throughout our conversations it became apparent that current interventions into government AI use were subject to a few recurring loopholes. One such loophole is the value threshold at which government procurement rules kick in. A typical procurement process has internal checks: A procuring agency must issue a request for proposals, evaluate the vendors that submit proposed contracts according to set metrics, subject contracts to legal vetting and, in the case of jurisdictions with AI governance frameworks, complete additional evaluations. Contracts below the value threshold, on the other hand, are rarely subject to any of these procedural checks.

However, even “small” AI purchases can have large reach and impact. In fact, pilot initiatives meant to test new technology for governments are, by definition, conducted on a scale that can easily fall below value thresholds.  For example, one city we spoke with had launched a parking study in a historically over-policed neighborhood. The study was through a vendor that recorded parking activity with cameras and analyzed it using AI. Since the project targeted only one city square (albeit a central one), it was under the value threshold and not subject to the full procurement process.

Another common loophole comes up when a nominally non-AI procurement process ends up including AI components. One official we spoke with described a welfare case management system that had escaped scrutiny in this way. The product’s goal was to assist a human caseworker in evaluating individual claims. The triaging system that determined the case order happened to be AI-driven (rather than by some other feature, such as order of receipt). This AI element, which could affect claim adjudication, was not contemplated at the start of the procurement process, and current procurement processes are not iterative. They proceed linearly. If an agency sets out to procure a product without contemplating AI as a possible solution, AI-specific provisions are not triggered—even if an AI system ends up playing a central role in the procured product.

More generally, mandatory programs around AI procurement are still rather new, and officials have told us that some agencies are simply unaware that they must comply with AI procurement checklists—or that such checklists even exist. This is further complicated in jurisdictions where only some types of AI products are subjected to mandatory reporting or evaluation, like the one proposed in the Executive Order. When a framework subjects only “high risk” AI applications to mandatory oversight, a procuring department often does not know whether a system is sufficiently risky for additional oversight. One official in charge of AI procurement told us this was the most frequent question their office received.

Finally, AI designed in house is not subject to procurement constraints. In jurisdictions with developed technical expertise, it is often cheaper to design risk assessments tools and other AI products in-house. Such a move avoids the procurement process entirely. This can eliminate problems that arise from an expertise mismatch between government agency and vendor. However, it removes procurement as a point of intervention and circumvents any discursive deliberation a procurement checklist might induce. According to one government attorney we consulted, this was a significant obstacle in the deployment of the WEF checklist in Brazil, where the majority of AI govtech products are developed in-house.

While these loopholes exist, AI checklists cannot fulfill either their competition or value based goals. Low-cost, hidden, and in-house AIs often exhibit the same risks posed by the systems that checklists do reach, namely high-cost and AI-specific systems. Like the welfare tool described above, they can have profound impacts on people’s lives and should be tracked to ensure continued alignment with policy goals. \\

\section{Substantive and Procedural Transparency are  Necessary\\ for Deploying Effective and Ethical AI systems}

Unlike more traditional procurement targets, e.g. vehicles or even accounting software, AI systems can have highly heterogeneous impacts.  They may work well most of the time, only to fail systematically for some small subset of use cases.  This heterogeneity makes vetting modern AI systems for compliance challenging. Even experts make mistakes and opaque internal vetting can leave problems in the dark.  While implemented assessments are comprehensive as to what should be vetted, they are less explicit about what kind of information should be made available to the public. When we spoke to creators and implementers of AI checklists, we found that in practice relatively little information was being shared. Specifically, additional transparency is needed in two different areas:
 
\emph{Substantive transparency} in government AI systems is important. We found in our discussions that while government officials were aware of expertise needs and potential procurement loopholes, they were less aware of how important public documentation processes were for identifying aspects that an expert may have missed. Sharing enough information for a diverse set of external stakeholders to meaningfully engage with system performance is a necessary condition for identifying those missing elements as well as refining our standards on how to evaluate AI systems.  (That is, once an issue is raised, it can be part of the standard checks that are performed on that system.) While some information must be kept private for security and privacy, as much information as possible about the training data, model architecture and performance, and post-deployment monitoring must be made available to the public. A registry of algorithms used in government, without such details, cannot let the general public assist in raising potential concerns about AI systems.

\emph{Procedural transparency} is often trivialized but is equally crucial. We found that even when certain procurement checks were technically required, agencies had a lot of leeway when it came to compliance. Many of the loopholes we found are examples of procedural, not substantive pitfalls. For example, above we noted that the application of some checklists depends on the risk level of the system, and the risk level depends on characteristics self-reported by the procuring agency. Internal incentives are not necessarily enough to outweigh the increased work and scrutiny of compliance.  
Working out the kinks of these reporting and monitoring systems and establishing best practices requires community consensus. By allowing additional parties insight into the AIs being procured, increased transparency can also make up for the lack of internal capacity for - and interest in - compliance.   In contrast, keeping the process internal and opaque will hobble progress towards standards.

\section{Building Towards Better Governance of Government AI}

Drawing on the observations in this piece, we conclude with a few systemic interventions that can help the current wave of legislation meaningfully manage public AI use:  

First, private-public-academic partnerships should center transparency and act synergistically. There is significant discussion regarding the need to expand AI expertise in government but there is little consensus on the appropriate path to accomplish this goal.  As we noted above, checklists and any content-based AI oversight framework will face scalability challenges. External experts---from civil society, industry and academia---will be needed to assist in shaping sector-specific regulatory requirements, identifying loopholes, and understanding AI limitations.  External experts will also likely be needed to conduct even the most specific checklist. 

Private firms are already providing their own versions of AI audits.  Just as how we have alignment on what a financial audit consists of, now is the time for public agencies creating and deploying these AI checklists to convene these private firms to create alignment on what an AI audit consists of.  Specifically: What criteria are vetted during an audit?  How, and in what detail?  What is the form of an audit report? This description of what makes an AI audit should be publicly released, building accountability in to an emerging market for the private sector AI audits.  By leading the way in clearly defining how governments will vet AI systems, the public sector can also inform standards for vetting AI systems in the private sector.

Second, liability must be apportioned thoughtfully. Alongside increased transparency, it is important that current and pending regulatory efforts define appropriate mechanisms for managing and apportioning liability. If excessive liability is placed on the government, agencies that could benefit from AI systems may choose not to adopt them---or, more concerningly, adopt them without the necessary process and transparency lest their disclosures be used against them in court. On the other hand, if facial compliance with a process is sufficient to avoid liability, it becomes an ineffective procedure for increasing safe and responsible use of AI systems. Moreover, agencies need guidance on apportioning liability with private partners, so that the govtech market does not become a place for corporate actors to offload risk to the public.

There is no way around the fact that AI systems are complex and require expert vetting. By recognizing the limits of substantive oversight and building towards transparent and expert-based processes for AI adoption, we begin to ensure that public AI use is aligned with public goals. We are hopeful that government efforts to engage AI experts in procurement processes, document good processes via checklists, and share lessons—good and ill—can trigger a race to the top around equity, access, and justice.\\

\textbf{Acknowledgements:} The authors would like to thank the officials within the Canadian, Singaporean, and Brazilian governments who made time to share their candid expertise on AI and the procurement process within their respective jurisdictions. We would also like to thank Kasia Chmielinski and Lauren Chambers for their support in developing the network necessary for this study, and Logan Fahrenkopf for helpful discussions and comments on the draft.   

\bibliography{long-version}

\begin{thebibliography}{10}
\providecommand{\natexlab}[1]{#1}
\providecommand{\url}[1]{\texttt{#1}}
\expandafter\ifx\csname urlstyle\endcsname\relax
  \providecommand{\doi}[1]{doi: #1}\else
  \providecommand{\doi}{doi: \begingroup \urlstyle{rm}\Url}\fi

\bibitem[Cecco(2020)]{The_Guardian_2020}
Leyland Cecco.
\newblock Google {S}idewalk {L}abs {T}oronto smart-city abondoned.
\newblock \emph{The Guardian}, May 2020.

\bibitem[Williams(2023)]{AIRegula99:online}
Zach Williams.
\newblock {AI} regulation ramping up in states from texas to connecticut.
\newblock \emph{Bloomberg Law}, August 2023.
\newblock (Accessed on 04/18/2024).

\bibitem[Con(2023)]{Connecti65:online}
Connecticut governor signs bill on {AI}, privacy; lawmakers consider {AI} policy.
\newblock \emph{IAPP}, June 2023.
\newblock (Accessed on 04/18/2024).

\bibitem[Folks(2024)]{Checking31:online}
Andrew Folks.
\newblock Checking in on proposed {C}alifornia privacy and {A}{I} legislation.
\newblock \url{https://iapp.org/news/a/checking-in-on-proposed-california-privacy-and-ai-legislation/}, March 2024.
\newblock (Accessed on 04/18/2024).

\bibitem[Malugade et~al.(2024)Malugade, Dullea, and Stauss]{2024AISt34:online}
Laura Malugade, Erik Dullea, and David Stauss.
\newblock 2024 {AI} state law tracker.
\newblock \url{https://www.huschblackwell.com/2024-ai-state-law-tracker}, March 2024.
\newblock (Accessed on 04/18/2024).

\bibitem[Exe(2023)]{Executiv42:online}
Executive {O}rder on the {S}afe, {S}ecure, and {T}rustworthy {D}evelopment and {U}se of {A}rtificial {I}ntelligence.
\newblock \url{https://perma.cc/7KDB-5NW8}, October 2023.
\newblock (Accessed on 04/18/2024).

\bibitem[Moss et~al.(2021)Moss, Watkins, Singh, Elish, and Metcalf]{DataSoci11:online}
Emanuel Moss, Elizabeth~Anne Watkins, Ranjit Singh, Madeleine~Clare Elish, and Jacob Metcalf.
\newblock Assembling accountability: Algorithmic impact assessment for the public interest.
\newblock \emph{Data \& Society}, June 2021.
\newblock (Accessed on 04/18/2024).

\bibitem[che(2021)]{checklists}
Algorithmic accountability for the public sector.
\newblock \emph{Ada Lovelace Institute, AI Now Institute and Open Government Partnership}, August 2021.
\newblock (Accessed on 04/18/2024).

\bibitem[CAD(2023)]{CADM}
Directive on automated decision-making- canada.ca.
\newblock \url{https://www.tbs-sct.canada.ca/pol/doc-eng.aspx?id=32592}, April 2023.
\newblock (Accessed on 04/19/2024).

\bibitem[Forum(2020)]{AIProcur11:online}
World~Economic Forum.
\newblock A{I} {P}rocurement in a {B}ox.
\newblock \url{https://www.weforum.org/publications/ai-procurement-in-a-box/}, June 2020.
\newblock (Accessed on 04/19/2024).

\end{thebibliography}

\end{document}